\documentclass{iopart}
\usepackage{graphicx}
\usepackage{amssymb}
\usepackage{color}
\pdfoutput=1
\newcommand{\text}[1]{\hbox{\scriptsize\rm #1}}
\newcommand{\PRD}{{\it Phys. Rev.} D}
\newcommand{\ApJ}{{\it Astrophys. J.}}

\begin{document}
\title[LIGO Binary Inspiral Search]
{Searching for Gravitational Waves from Binary Inspirals with LIGO}
\author{Duncan~A.~Brown$^1$,
Stanislav~Babak$^2$,
Patrick~R.~Brady$^1$,
Nelson~Christensen$^3$,
Thomas~Cokelaer$^2$,
Jolien~D.~E.~Creighton$^1$,
Stephen~Fairhurst$^1$,
Gabriela~Gonz\'{a}lez$^4$,
Eirini~Messaritaki$^1$,
B.~S.~Sathyaprakash$^2$,
Peter~Shawhan$^5$, and
Natalia~Zotov$^6$
}
\address{$^1$ University of Wisconsin--Milwaukee, Milwaukee, WI 53201, USA\\
  $^2$ Cardiff University, Cardiff, CF2 3YB, United Kingdom\\
  $^3$ Carleton College Northfield, MN 55057, USA\\
  $^4$ Louisiana State University, Baton Rouge, LA 70803, USA\\
  $^5$ California Institute of Technology, Pasadena, CA 91125, USA\\
  $^6$ Louisiana Tech University, Ruston, LA 71271, USA}
\begin{abstract}
  We describe the current status of the search for gravitational waves from
  inspiralling compact binary systems in LIGO data. We review the result from
  the first scientific run of LIGO (S1). We present the goals of the search of
  data taken in the second scientific run (S2) and describe the differences
  between the methods used in S1 and S2.
\end{abstract}
\ead{duncan@gravity.phys.uwm.edu}

\section{Introduction}
\label{s:intro}

The Laser Interferometer Gravitational Wave Observatory
(LIGO)\cite{Barrish:1999} has completed three science data taking runs. The
first, referred to as S1, lasted for 17 days between August 23 and September
9, 2002\cite{abbott2003a}; the second, S2, lasted for 59 days between February
14 and April 14, 2003; the third, S3, lasted for 70 days between October 31,
2003 and January 9, 2004.  During the runs, all three LIGO detectors were
operated: two detectors at the LIGO Hanford observatory (LHO) and one at the
LIGO Livingston observatory (LLO).  The GEO detector in Hannover, Germany
operated during S1 and in the latter part of S3 from December 30 2003 to the
end of the run. The detectors are not yet at their design sensitivity, however
the detector sensitivity and amount of usable data has improved between each
data taking run. The noise level is low enough that searches for coalescing
compact binaries are worthwhile and since the start of S2, these searches are
sensitive to extra-galactic sources.

The analysis of the LIGO data for gravitational waves from coalescing neutron
star binaries has been completed for S1\cite{abbott2003b} and
S2\cite{abbott2004a}, and is in progress for S3. Additional searches for
binary black hole coalescence and binary black hole MACHOs in the galactic
halo are underway using the S2 and S3 data. Here we review the result of the
S1 search and describe the scientific goals of the searches of the S2 data. We
review the S2 binary neutron star search and highlight the differences between
the methods used in S1 and those currently employed.

\section{Results from the first LIGO Science Run}
\label{s:s1}
The S1 analysis for inspiralling neutron star binaries searched a total of 236
hours of LIGO data from the 17 day run. Data from the GEO
detector and the Hanford 2km (H2) interferometer were not used in this
analysis, since the sensitivity of these instruments was significantly less
than the Hanford 4km (H1) and Livingston 4km (L1) interferometers. The
additional data from these instruments would not have provided increased
confidence in a detection or significantly decreased the upper limit on the
rate. The amount of \emph{double coincident data}, defined as data taken when
both L1 and H1 were operating, was only 116 hours during S1. The decision was
therefore taken to also use \emph{single interferometer data}, data taken when
only one of the L1 or H1 interferometers was operating, to produce the upper
limit. This meant that there were times in the analysis when a candidate event
could not have been confirmed by coincidence, but the amount of data available
for the upper limit was increased to 236 hours.

The distance to which an interferometric detector is sensitive to
gravitational radiation from a coalescing binary depends on the noise spectrum
of the interferometer and the component masses of the binary system. To
provide a standard measure of the range of a detector we chose the distance to
which we can detect a pair of non-spinning \emph{optimally oriented} $1.4
M_{\odot}$ neutron stars at a signal-to-noise ratio $\rho = 8$. An optimally
oriented binary is located directly above the $z$-axis of the detector, with
the arms of the detector defining the $x$ and $y$-axes, and the angular
momentum vector of the binary system is oriented parallel to the $z$-axis of
the detector. In S1 the maximum distance of L1 was $176$~kpc and the maximum
distance of H1 was $46$~kpc. The interferometers were sensitive to inspirals
in the Milky Way and the Magellanic Clouds.

No double coincident candidates were found in the S1 data. The loudest
inspiral trigger found had a signal-to-noise ratio $\rho = 15.9$ in L1
detector. Further analysis of this event showed that it was due to a
photodiode saturation and not a gravitational wave. In addition the next nine
loudest events from the interferometers were also examined. We investigated
the behavior of the signal-to-noise and $\chi^2$ near the inspiral trigger. 
We also examined the time frequency structure of the interferometer data
associated with the trigger. None of the triggers examined were consistent
with an binary neutron star inspiral signal of astrophysical origin; the
triggers were consistent with instrumental misbehavior. The analysis of these
false triggers has suggested better tests of data quality prior to the
analysis\cite{gwdawveto} used in S2 as well as additional signal based veto
methods\cite{gwdawpeter}, which will be used in future analyses.

The S1 binary neutron star search set an upper limit of $\mathrm{R}_{90\%} <
1.7 \times 10^2$ per year per Milky Way Equivalent Galaxy (MWEG) with no
gravitational wave signals detected. Details of this analysis can be found in
\cite{abbott2003b}

\section{Analysis Goals for the Second LIGO Science Run}
\label{s:s2goals}

\begin{figure}[tbh]
  \vspace{5pt}
  \begin{flushright}
    \includegraphics[width=\textwidth]{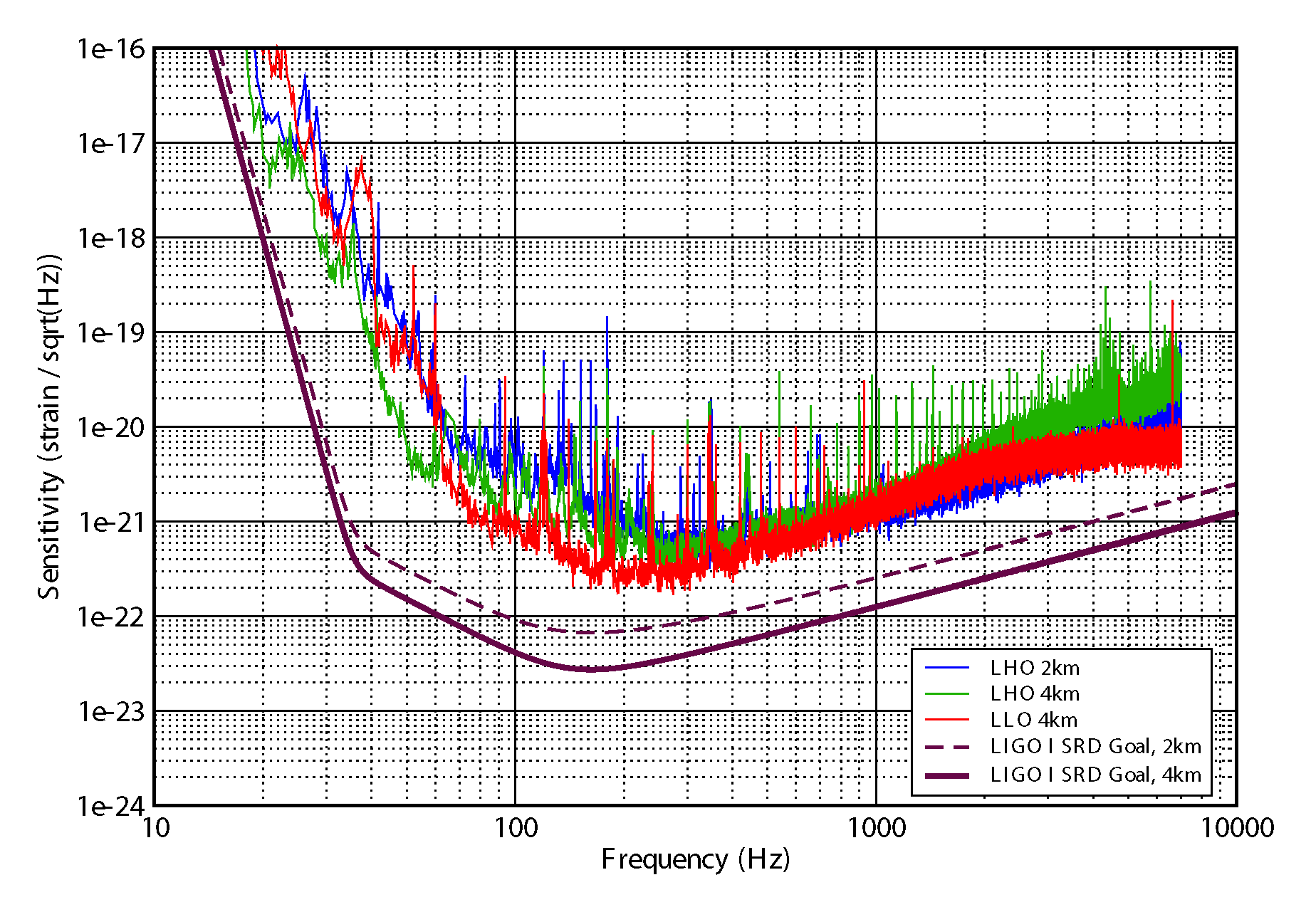}    
  \end{flushright}
  \caption{%
  Typical sensitivities of the three LIGO interferometers during the second
  LIGO science run\cite{s2noisecurve} shown as strain amplitude spectral
  density, $\tilde{h}/\sqrt{\mathrm{Hz}}$. The smooth solid curve shows the
  design sensitivity (SRD Goal) of the $4$~km interferometers and the smooth
  dashed curve shows the design sensitivity of the $2$~km interferometer.
  }
\label{f:s2noisecurve}
\end{figure}

Figure~\ref{f:s2noisecurve} shows the typical sensitivity of the LIGO
interferometers during S2. Both the average range and the duty cycle of the
interferometers has improved since S1. The average distance to the optimally
oriented binary described in section~\ref{s:s1} is $1.81$~Mpc for L1,
$0.90$~Mpc for H1 and $0.60$~Mpc for H2. This increase in range means that the
detectors are now sensitive to binary inspirals in Andromeda as well the Milky
Way and Magellanic Clouds.  There is also some sensitivity to the galaxies
M33, M32 and M110. This is a significant improvement over S1. A further
increase in sensitivity and population is expected in the S3 data.

The increase in the amount of time when the data was suitable for analysis has
allowed us to restrict the S2 analysis to times of double and triple
coincident data. We are pursuing three different searches in the S2 data.

\subsection{Binary Neutron Stars}
\label{ss:bnssearch}

We continue to search for gravitational waves from inspiralling binary neutron
stars with component masses in the range $1$ to $3 M_\odot$.   The scientific
goal of the data analysis in S2 is the detection of a gravitational wave from
a coalescing binary neutron star system. In the absence of a detection, the
population models available allow us to place upper limits on the rate of
inspiralling compact binaries in the universe.  The non-spinning binary systems
that we search for have well modeled waveforms using the second order
post-Newtonian approximation\cite{biww}. It is expected that spin does not
significantly decrease the detection efficiency of these systems. We describe
this search in more detail in section~\ref{s:bnspipeline} below.

\subsection{Non-Spinning Binary Black Holes}
\label{ss:bbhsearch}

As mentioned above, the sensitivity of the inspiral search depends on the
masses of the objects in the binary system. For binary system of component
masses $m_1$ and $m_2$, the signal-to-noise ratio scales with the \emph{chirp
mass}, $\mathcal{M} = M\eta^{3/5}$. Here $M = m_1 + m_2$ is the total mass of
the binary and $\eta = m_1 m_2/M^2$. The scaling is approximately $\rho
\propto \mathcal{M}^{5/6} / d$ where $d$ is the distance to the
binary\cite{300yrs}.  This means that if an the optimally oriented binary
described above consisted of a pair of $10M_\odot$ black holes then it would
produce signal-noise ratio $8$ at a distance of $9.32$~Mpc, rather than at a
distance of $1.81$~Mpc for a pair of $1.4 M_\odot$ neutron stars. We therefore
wish to pursue a search for gravitational waves from inspiralling binary black
holes.  For systems with a total solar mass above $6\ M_\odot$ the waveforms
used in the binary neutron star search become unreliable. The post-Newtonian
expansion begins to break down as the orbital velocity becomes relativistic.
Several possible waveforms have been suggested for detection of gravitational
waves from binary black holes\cite{damouriyersathya}.  Bounnano, Chen and
Valisineri (BCV) have proposed a \emph{detection template family} (DTF) of
waveforms\cite{bcv} that can be used to search for several classes of binary
black hole waveforms in a single search.  The goal for S2 is a to use the
non-spinning BCV DTF waveforms to search for the gravitational radiation from
binary black holes.  The effect of spin can be significant for binary black
hole systems, however here we restrict ourselves to non-spinning binaries for
simplicity. A search for spinning black hole binaries is in the early stages
of development.

\subsection{Binary Black Hole MACHOs}
\label{ss:machosearch}

Observations of gravitational microlensing of stars in the Magellanic clouds
suggest that between $8\%$ and $20\%$ of the galactic halo is composed of a 
population of massive astrophysical compact halo objects (MACHOs) of mass
between $0.15$ and $0.9 M_\odot$\cite{alcock2000}. It has been suggested that
if these MACHOs are primordial black holes (PBHMACHOs) then some fraction
of the PBHMACHOs may be in binary systems and could be detectable by ground
based interferometers such as LIGO\cite{nstt}. Although these binaries
are a speculative source, modeling of their formation in the early universe
suggests a detection rate significantly higher than that of binary neutron
stars\cite{ioka}. The second order post-Newtonian waveforms used in the
binary neutron star search provide excellent template for these PBHMACHO
binaries; the low mass of these systems means that LIGO is sensitive to a much
earlier stage of inspiral when the orbital velocity is low. We will search for
these systems and, in the absence of detection, place an upper limit on the
rate using population models obtained from galactic halo densities.

\section{S2 Binary Neutron Star Search}
\label{s:bnspipeline}

We have made several modifications to the binary neutron star data analysis
pipeline since the S1 analysis. In the S2 analysis we consider only coincident
interferometer data; a total of $355$ hours of data has been used for the
analysis. This consists of $231$ hours of triple coincident data when all
three LIGO interferometers were operating, $92$ hours of data where only L1
and H1 were operating and 31 hours when only L1 and H2 were operating. We
select a subset (approximately $10\%$) of the data to be used as
\emph{playground}. This is used to tune the various thresholds in the analysis
pipeline. To ensure that the playground is representative of the entire run,
we select $600$~sec of data every $6370$~sec as playground data\cite{T030256}.
In the absence of a detection, we can set an upper limit on the rate of
inspirals in the universe using data not in the playground\cite{jolien}.
Excluding the playground ensures that we do not introduce a statistical bias
into the upper limit by tuning on data that is used to produce the limit. We
do not exclude the possibility of a detection in the playground data.

In order to avoid the possibility of correlated noise sources affecting the
background rate estimation, we chose not to use the data when only the H1 and
H2 interferometers were in operation. This data, as well as the discarded single
interferometer data, will be used in coincidence with data from the TAMA
detector in a search for galactic inspiralling binaries. In this section we
describe the methods employed in the S2 binary neutron star search.

\subsection{Inspiral Trigger Generation}
\label{ss:trigger}

We search for inspiral signals in the LIGO data with a matched
filtering\cite{wz} algorithm implemented in the \emph{findchirp}
package\cite{findchirp} of the LIGO/LSC Algorithm library\cite{lal}. The LIGO
data is recorded at a sampling rate of $16384$ Hz.  The highest frequency of
gravitational radiation that we are searching for is approximately $2200$ Hz and
so we resample the data to $4096$ Hz for the matched filtering.  An 8th order
Butterworth filter was applied to the interferometer data which attenuated the
signal by $10\%$ at $100$ Hz. This prevented numerical corruption of the power
spectral estimate due to large power in the LIGO noise curve at low
frequencies. Initial analysis of the data from the L1 interferometer
discovered that a large non-stationary noise source at around $60$---$70$~Hz
was producing an excessive number of inspiral triggers. A low frequency cutoff
was applied in the frequency domain by setting the data to zero at frequencies
below $100$ Hz.  The shape of the noise power spectra was such that this did
not produce a significant loss in inspiral range. 

Data is analyzed in $2048$~sec \emph{analysis chunks} consisting of fifteen
$256$~sec \emph{analysis segments} which are overlapped by $128$~sec. A median
power spectrum is computed for each of these analysis segments; for each
frequency bin, the median value of the $15$ power spectra is used to
calculate the average power spectrum used in the matched filter.  A
\emph{template parameter bank} is used to generate second order post-Newtonian
templates in the frequency domain using the stationary phase approximation to
the inspiral signal. For a given template
and analysis segment we construct the signal-to-noise ratio, $\rho$,  and
search for times when this exceeds a threshold,  $\rho > \rho^\ast$. If this
happens, we construct a template based veto, the $\chi^2$
veto\cite{brucechisq}. Small values of $\chi^2$ indicate that the
signal-to-noise was accumulated in a manner consistent with an inspiral
signal. If the value of the $\chi^2$ veto is below a threshold, $\chi^2 <
{\chi^2}^\ast$, then an inspiral trigger is recorded at the maximum value of
$\rho$. For a given template multiple triggers can be recorded in a segment.
The triggers are clustered so that distinct triggers are separated by at least
the length of the template.  Each analysis segment is filtered through all the
templates. It is possible for multiple templates to trigger at same time.
Details of the template banks passed to the trigger generation code are
described in the next section. 

\subsection{Data Analysis Pipeline}
\label{ss:pipeline}

The interferometer operators, in consultation with scientific monitors present
at the observatory during data taking, flag times when the interferometers are
in stable operation and the data is suitable for analysis.  Further studies of
the raw data yield a series of \emph{data quality cuts} that are used to
exclude anomalous data from the inspiral analysis\cite{gwdawveto}. We have
excluded times when \emph{(i)} servo controls in the L1 interferometer were
set incorrectly, \emph{(ii)} calibration information is unavailable for the
analysis, \emph{(iii)} there are photodiode saturations, \emph{(iv)} data has
invalid time stamp information and \emph{(v)} the noise in the H1
interferometer is significantly larger than average. In general, the
interferometer is considered to be malfunctioning during these times with the
exception of \emph{(i)} and \emph{(ii)} which are due to operator error. In
the case of \emph{(v)}, we ensure that the increased noise is not due to the
presence of an inspiral signal in the data by only excluding times when the
noise is excessive for more than $180$~sec, which is significantly longer that
our longest inspiral signal of $3.7$~sec.

As can be seen from figure~\ref{f:s2noisecurve} and the average sensitivity
during S2, the range of the L1 detector is approximately twice that of the H1
detector, which is larger than that of the H2 detectors. At all times during
the run L1 is more sensitive than H1 and H1 is more sensitive than H2. We use
this and the fact that we demand that a trigger be present in multiple
interferometers to construct a \emph{triggered search pipeline}. This allows
us to save a significant amount of computational effort during the search,
without reducing the detection efficiency. Here we illustrate the method of
the triggered search for two interferometers. Further details of the triggered
search pipeline for multiple interferometers can be found in
reference~\cite{abbott2004a}.

For each analysis chunk a template bank is generated for the L1 detector for
binary neutron stars with component masses between $1.0$ and $3.0 M_\odot$, as
described in \cite{owensathya}.  The \emph{minimal match} of the bank is
chosen to be $0.97$. A random inspiral signal lying in the space of the bank
would lose no more than $3\%$ of the signal-to-noise ratio due to mismatch
between the signal and the nearest template. We filter each L1 analysis chunk
through the corresponding bank to generate inspiral triggers. We then select
each template from the L1 bank that produced one or more inspiral triggers to
construct a \emph{triggered bank.} The triggered bank, which is a subset of
the original template bank, is used to filter the data from the less sensitive
interferometer (e.g. the H1 detector) to produce a second list of inspiral
triggers. We demand coincidence between triggers from the different
interferometers, as described below, to produce the list of \emph{coincident
triggers.} We apply instrumental vetoes to this list of coincident triggers to
exclude triggers that are due to known instrumental or environmental artifacts
in the data, as described in \cite{gwdawveto}. Any surviving triggers are
considered to be the list of candidate inspiral triggers from the analysis.

To perform the triggered search pipeline on the full data set, we constructed
a directed acyclic graph (DAG) that described the work flow.  The DAG is
executed using the Condor high throughput computing system\cite{condor} on the
UWM and Caltech  clusters.

\subsection{Trigger Coincidence}
\label{ss:coincidence}

For a trigger to be considered coincident in two interferometers, we demand
that it is observed in both interferometers within a temporal coincidence
window $\delta t$. Monte Carlo analysis with simulated signals suggests that we
can measure the time of a trigger to within $1$~ms, so
we demand $\delta t = 1$~ms if the interferometers are located at the same
observatory. If the detectors are not co-located, we allow for the $10$~ms
light travel time between the LIGO observatories by demanding $\delta t =
11$~ms. We also demand that the waveform of the triggers are consistent by
requiring that the two mass parameters, $m_1$ and $m_2$, of the binary are
identical.

\begin{figure}[tbh]
  \vspace{5pt}
  \begin{flushright}
    \includegraphics[width=\textwidth]{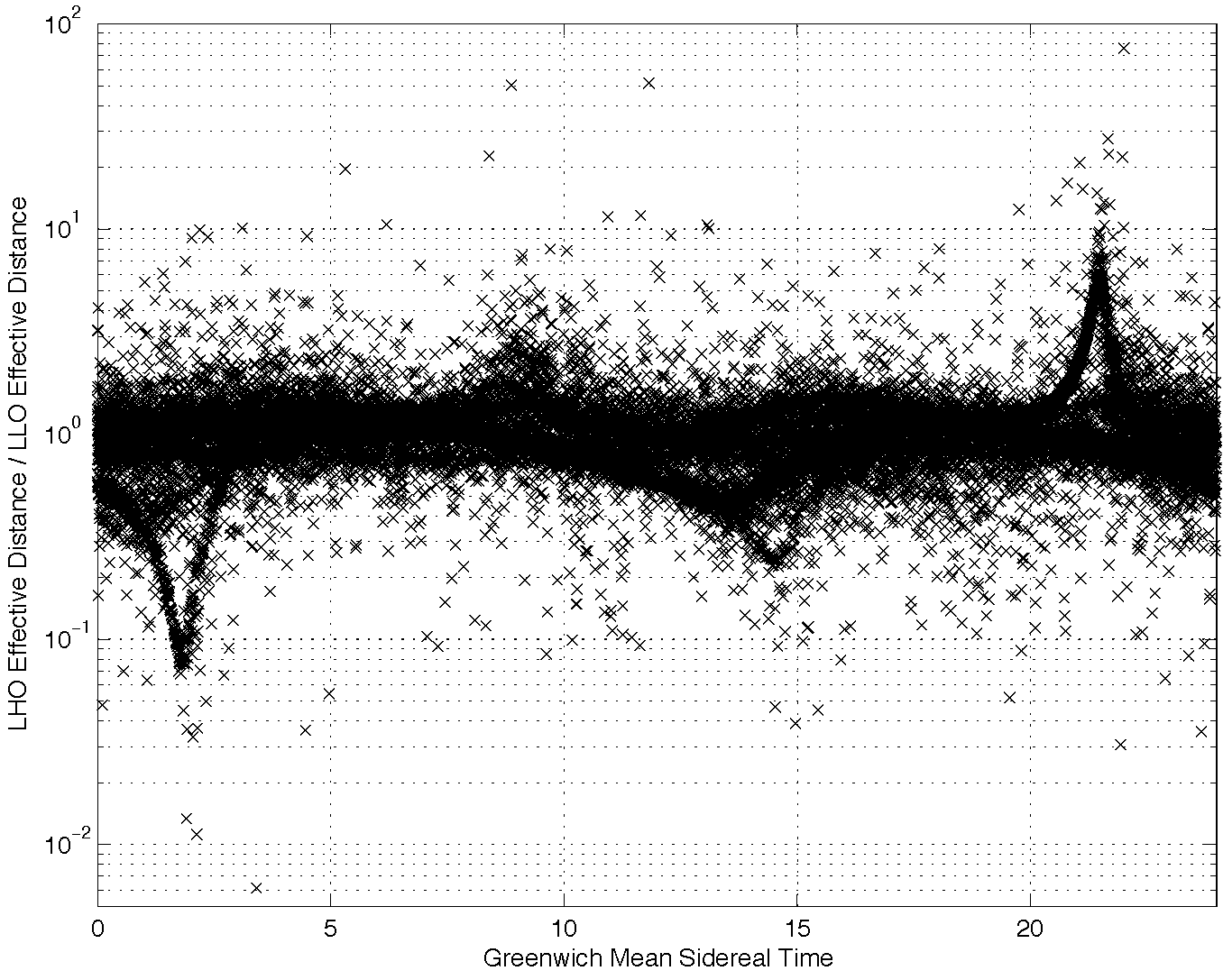}    
  \end{flushright}
  \caption{%
  The ratio of the known effective distance of an injected signal in the
  Hanford Observatory (LHO) to the known effective distance of an injected
  signal in the Livingston Observatory (LLO) as a function of Greenwich Mean
  Sidereal Time. The slight misalignment of the interferometers at the two
  different observatories due to the curvature of the earth causes the antenna
  pattern of the detectors to differ. As a result the distance at which a
  binary system appears is different in each detector, even in the absence of
  noise.  The ratio of effective distances can be significant, so this
  precludes the use of an amplitude cut when testing for inspiral trigger
  coincidence between different observatories.
  }
\label{f:gmst_dist_ratio}
\end{figure}
We now consider an amplitude cut on the signals. The Livingston and Hanford
detectors are not co-aligned. There is a slight misalignment of the detectors
due to the curvature of the earth and so the antenna patterns of the detectors
differ. This causes the measured amplitude of a gravitational wave to differ
between the sites. In the extreme case, it is possible for a binary to be
completely undetectable by the L1 detector, but still detectable by the H1 and
H2 detectors. For a given inspiral trigger, we measure the \emph{effective
distance} of the binary system. This is the distance at which an optimally
oriented binary would produce the observed signal-to-noise ratio.
Figure~\ref{f:gmst_dist_ratio} shows the ratio of effective distances between
the two LIGO observatories for the population of binary neutron stars
considered in the S2 analysis. The significant variation of the effective
distance precludes using a naive test for amplitude coincidence. It is
possible to obtain information about sky position from time delay between
sites to construct a more complicated amplitude cut, but this has not be used
in the S2 analysis.

In the case of triggers from the H1 and H2 interferometers that are coincident
in time and mass, we apply an amplitude cut that tests that the effective
distance of the triggers is coincident given the relative sensitivity of the
detectors, while allowing for error in this measurement which is determined by
Monte Carlo simulations.  When testing for triple coincident triggers we 
accept triggers that are coincident in the L1 and H1 detectors that are
\emph{not} present in the H2 detector \emph{if} the effective distance of the
trigger is further than the maximum distance to which H2 is sensitive at 
the time of the candidate trigger. This maximum distance is dependent on the
both the sensitivity of H2 at the time of the candidate trigger and the
signal-to-noise threshold, $\rho^\ast$, chosen for H2.

As in the S1 analysis, the list of surviving candidate triggers is followed up
by examining the raw gravitational wave data, auxiliary interferometer channels
and physical environment monitoring channels to determine if the triggers are
truly of astrophysical origin.

\subsection{Background Estimation}
\label{ss:background}

Since we restrict the S2 analysis to coincident data and require that at least
two of the interferometers must be located at different observatories, we may
measure a background rate for our analysis. After generating triggers for each
interferometer, we slide the triggers from one observatory relative to the
other observatory and look for coincidences between the shifted and unshifted
triggers. The minimum slide length is chosen to be greater than the length of
the longest filter ($20$~sec) so any coincident triggers detected must be due to background and
not astrophysical events. By examining the distribution of background events
in the $(\rho_\mathrm{H},\rho_\mathrm{L})$ plane we can attempt to determine
contours of constant false alarm rate in order to construct a combined
effective signal-to-noise ratio for a coincident trigger\cite{abbott2004a}.

\subsection{Detection Efficiency}
\label{ss:eff}

In absence of detection, we will construct an upper limit on event rate.  To
do this, we must measure the detection efficiency of the analysis pipeline to
our population. A Monte Carlo method is used to measure this efficiency. We
simulate a population of binary neutron stars\cite{nutzman} and \emph{inject}
signals from that population into the data from all three LIGO
interferometers. The injection is performed in software by generating an
inspiral waveform and adding it to interferometer data immediately after the
raw data is read from disk. We inject the actual waveform that would be
detected in a given interferometer accounting for both the masses,
orientation, polarization, sky position and distance of the binary, the
antenna pattern and calibration of the interferometer into which this signal
is injected.  The effectiveness of software injections for measuring the
response of the instrument to an inspiral signal is validated against
\emph{hardware injections}\cite{hw} where an inspiral signal is added to the
interferometer control servo during operation to produce the same output
signal as a real gravitational wave.  The data with injections is run through
the full analysis pipeline to produce a list of inspiral triggers. The
detection efficiency of the pipeline, $\epsilon$, is the ratio of the number
of detected signals to the number of injected signals.

\section{Conclusion}
\label{s:conclusion}

The S1 binary neutron star search is now complete\cite{abbott2003b}. No
coincident gravitational wave candidates were found and an upper limit of
$R_{90\%} < 1.7 \times 10^2$ per year per MWEG was set on the rate of
inspiralling binary neutron stars. Results of the S2 binary neutron star search
are currently under LSC review and will be available shortly. In addition to
this we will soon be in a position to present results from the non-spinning
binary black hole search and the search for binary black hole MACHOs in the
galactic halo.

\ack
The authors gratefully acknowledge the LIGO Scientific Collaboration, who made
the LIGO science runs possible. This work was supported by grants from the
National Science Foundation, including grants 
PHY-0200852, 
PHY-0244357, 
PHY-0135389, PHY-0355289, 
and PHY-0107417, 
and by Particle Physics and Astronomy Research Council grant PPA/G/O/2001/00485.
PRB is grateful to the Alfred P. Sloan Foundation and the Research
Corporation Cottrell Scholars Program for support.

\end{document}